\documentclass[prb,aps,showpacs,twocolumn,superscriptaddress,10pt,amsmath,amssymb,nofootinbib]{revtex4-1}
\usepackage{epsfig}
\usepackage{amsmath}
\usepackage{amssymb}

\newcommand{\bise}{Bi$_2$Se$_3$}
\newcommand*{\bb}[1]{\boldsymbol{#1}}

\makeatletter 

\begin{document}

\title{\it  Gap and spin texture engineering of Dirac topological states at the Cr--\bise\ interface} 

\author{H. Aramberri}
\affiliation{Instituto de Ciencia de Materiales de Madrid, ICMM-CSIC, Cantoblanco, 28049 Madrid, Spain.}

\author{M.C. Mu\~noz}
\affiliation{Instituto de Ciencia de Materiales de Madrid, ICMM-CSIC, Cantoblanco, 28049 Madrid, Spain.}

\date{\today}
\begin{abstract}
The presence of an exchange field in topological insulators reveals novel spin related phenomena derived
from the combination of topology and magnetism. In the present work we show 
the controlled occurrence of either metallic or gapped topological Dirac states
at the interface between ultrathin Cr films and the
\bise\ surface. The opening and closing of the gap at the Dirac point 
is caused by the spin reorientation transitions arising in
the Cr films.
 We find that atom-thin layers of Cr adhered to \bise\ surfaces 
present a magnetic ground state with ferromagnetic planes coupled 
antiferromagnetically. As the thickness of the Cr film increases
stepwise from one to three atomic layers, the direction of the magnetization
changes twice from out--of--plane to in--plane and to out--of--plane again. 
The out--of--plane magnetization drives the gap opening and the topological surface states acquire a
circular meron spin structure. Therefore, the Cr 
spin reorientation leads to the metal-insulator transition in the \bise\ 
surface and to the correlated modification of the surface state spin texture. 
Consequently, the thickness of the Cr film provides an effective and 
controllable mechanism to modify the metallic or gapped nature, as well as the spin texture of 
the topological Dirac states.
\end{abstract}

\maketitle

\section{Introduction}
The recent discovery of three dimensional (3D) topological insulators (TIs)
has led to unique and fascinating physical phenomena, such as the quantum anomalous Hall (QAH) phase~\cite{QAHEtheo}
 and the topological magnetoelectric effect~\cite{TME1}.
The robustness of the surface metallicity under time 
reversal invariant perturbations and the realization of novel quantized 
states arising from their peculiar coupling to magnetic fields
are distinct characteristics of this new phase of quantum matter. 
A central feature of TIs is the existence of helical surface states (SSs) with
the electron spin locked to the crystal momentum~\cite{zhang2009topological,zhang2010njop}.
The presence in the TI of an exchange field, which violates
time reversal symmetry (TRS), lifts the Kramers 
degeneracy and discloses novel spin related phenomena 
directly derived from the combination of topology and magnetism.
The QAH effect has already been observed in 
three-dimensional magnetic TIs~\cite{QAHEexp}. Nevertheless, the experimental 
realization of a magnetoelectric topological insulator, which in fact 
corresponds to a {\it non--integer} quantum Hall effect 
at the surface~\cite{QAHEtheo,TME2}, still remains a challenge.

To experimentally achieve these topological phases a surface gapped by a
TRS breaking perturbation is required. 
There are three different ways to 
break TRS in TIs,
 either by conventional doping with magnetic elements~\cite{jap_Cr,biteMn,nanoCr2}, by  
proximity to a magnetic film at a TI-magnetic interface~\cite{qin2014,liu2015}, or
 by an external magnetic field~\cite{TIsmagB}. 
The effect of magnetic doping with 3$d$ transition metals in 
the \bise\ family of compounds has been extensively investigated both 
theoretically and experimentally~\cite{ChenCr,nanoCr2,APL_Cr,prlCr2,nanoCr3,jap_Cr,Crrelaxexp,jap_Cr2,nanoCr4}. 
It has been shown that the interaction with magnetic impurities
modifies the electronic and magnetic ground state of the 3D TIs. 
However, the changes in the ground state are not universal since they 
critically depend on the specific magnetic atoms, occupation sites of the 
magnetic impurities~\cite{APL_Cr}, and experimental conditions.
Cr-doped \bise\ is a prototype magnetic TI, and 
several works~\cite{ChenCr,nanoCr2,APL_Cr,prlCr2,nanoCr3,jap_Cr} have reported 
magnetically induced effects in this system. 
 First-principles calculations found that substitutional Cr, 
which is energetically more favorable than interstitial Cr,
 preserves the insulating character
in the bulk and that Cr-doped \bise\ is likely to be ferromagnetic~\cite{CrPRL,CrPRB}.
However, evidence from the experimental observations 
 is so far inconclusive~\cite{ChenCr,APL_Cr,jap_Cr,Crrelaxexp,jap_Cr2,nanoCr4}.
 Both ferro--~\cite{APL_Cr} and antiferromagnetism~\cite{jap_Cr} have been reported 
and, recently, the coexistence of both ferro-- and 
antiferromagnetic Cr defects in high quality epitaxial 
thin films has been observed~\cite{acsnanoCr}. Nevertheless,  
Cr doping of bulk or thin films of \bise\
crystals seems to lead to a gap opening in the Dirac cone, evidencing 
time--reversal symmetry breaking~\cite{APL_Cr,jap_Cr,ChenCr}. In contrast,  
 surface deposition of Cr atoms on the surface of \bise\ 
up to $\approx$~10\% monoatomic layer (ML) coverages 
 preserves the metallic surface~\cite{nanoCr}. The absence of gap 
opening at the Dirac point indicates that for dilute Cr adatom concentrations
there is no long--range out--of--plane ferromagnetic order. 
Despite these works, the interface between Cr films and the \bise\ surface has
not been investigated and thus the spin behavior of the topological SS under the interaction 
with ultrathin Cr films remains unknown.

In the present work we explore the spin configuration and topological state
at the interface of \bise\ surfaces and Cr films in the ultrathin limit, one to three MLs thick.
We find that the presence of the Cr magnetic film triggers a double transition,
 from a Dirac-metal to a gapped system, on the topological SS of \bise\ as a function of the Cr thickness.
The gap opening at the Dirac point is induced by the proximity 
of the Cr film and thus the observed modulation of the gap is associated with the 
spin reorientation occurring in the magnetic layer. In fact, the magnetization direction in the 
Cr film evolves from out--of--plane to in--plane and once again to out--of--plane as the 
Cr thickness increases stepwise from one to two and three MLs. 
Correlated with the gap, there is a modulation of the spin 
texture of the topological SSs, which undergoes a double circular skyrmion to circular meron
transition.
 
\section{Model and Methods}
Density Functional Theory (DFT) spin--polarized calculations were carried out with  
the SIESTA code~\cite{soler2002siesta} as implemented in the GREEN package~\cite{green,greensocjorge},
 although specific structures 
were also calculated with the Vienna \textit{ab-initio} simulation package (VASP) \cite{VASP-Kresse2-PhysRevB.48.13115}.
The generalized gradient approximation with Perdew-Burke-Ernzerhof~\cite{pbegga} type 
exchange-correlation functional was used in all cases.
In the SIESTA calculations, the spin--orbit coupling is considered via the
recently implemented fully-relativistic pseudopotential formalism~\cite{greensocjorge},
while the semi-empirical pair-potential approach to van der Waals
(vdW) forces of Ortmann \textit{et al.}~\cite{vdW-SIESTA} was employed to correctly account for the weak 
inter quintuple layer (QL) interaction in the \bise\ crystal. The numerical atomic orbitals basis set 
was generated according to the double $\zeta$-polarized scheme with confinement 
energies of 100 meV. For the computation of three-center integrals, a 
mesh cut-off as large as 1200 Ry was used, equivalent to a real space
grid resolution below 0.05 \AA$^3$.
 In the VASP calculations 
plane wave basis set with a kinetic energy cutoff of 340 eV was used. 
For the Brillouin zone (BZ) integrations a centered 13$\times$13$\times$1 k-sampling 
was employed, while the electron temperature was set to k$_B$T=10~meV in both calculation schemes.

\bise\ has a rhombohedral crystal structure with space group R$\bar{3}$m 
(D$^5_{3d}$). It can be described as a layered compound constituted by 
 QLs along the [0001] direction. A QL contains alternating Se and Bi atomic 
layers, and within each QL the two Bi layers are equivalent, while the Se in the 
middle is unequivalent to the external Se. The stacking pattern is  $fcc$--like, 
-AbCaB-CaBcA-, where capital and
small letters stand for Se and Bi, respectively. The Se-Bi bonds within 
the QLs are mainly covalent, while at adjacent QLs the Se-Se double-layer 
is only weakly bonded through van der Waals forces.
The in--plane lattice parameter is $a_{\mathrm{Bi}_2 \mathrm{Se}_3}$=4.14~\AA, while $c$=9.54~\AA\ 
determines the periodicity along the [0001] direction.

\begin{figure}
  \includegraphics[scale=0.038]{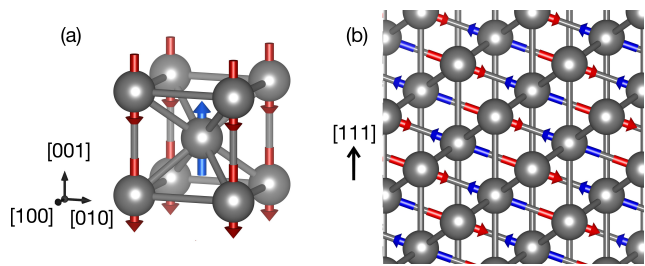}
  \caption{ (a) \textit{bcc} unit cell of bulk Cr. Arrows indicate the magnetic moment. The two unequivalent atoms    
            show opposite magnetic moment. (b) Side view of bulk Cr with the [111] direction as indicated in the figure.
            Cr exhibits an ABC stacking pattern along this direction, with opposite magnetic moments for alternating atomic planes.
            Two out of the eight first nearest neighbors of each atom lie in the first layer above and below, while the remaining six
            lie three layers above and below.}
  \label{bulkCr}
 \end{figure}

 Bulk Cr follows a \textit{bcc} crystal structure with lattice parameter $a_{\mathrm{Cr}}$=2.91~\AA. 
 Each atom has 8 nearest neighbors (n.n.s). Surfaces perpendicular to the [111] direction exhibit
 three-fold C$_{3}$ symmetry and 
 an open structure, since only 6 out of the 8 n.n.s lie in the adjacent atomic layers, while the remaining 
 2 n.n.s are located three atomic layers above and below. Along this direction the stacking sequence 
 follows an ...ABCABC... pattern, analogous to that of the  \bise\ crystal in the 
[0001] direction (see Figure ~\ref{bulkCr}).
Cr is unique among the 3$d$ transition metals, showing an itinerant 
antiferromagnetic ground state.
It exhibits a spin density (SDW) wave along
 the [100] direction --or, equivalently, along the [111] direction--, with a wave vector almost commensurate
 with the lattice, being its N\'eel temperature T$_N$ =311 K. 
Contrary to what happens in bulk crystals, in which all three 
crystallographic directions are equivalent, in thin Cr films the 
SDW wave vector is perpendicular to the film surface and the SDW is commensurate 
with the lattice. 
Since the Cr--\bise\ systems studied are formed of a maximum of 3 Cr layers, 
we can consider Cr as a pure antiferromagnet in the ultrathin film regime.
Thus, the Cr slabs are expected to show atoms in the 
same atomic layer coupled ferromagnetically, being the interlayer coupling  
antiferromagnetic.

We model the 
Cr--\bise\ interfaces by 1$\times$1$\times$1 and 2$\times$1$\times$1
 supercells with the equilibrium in--plane lattice constant of
bulk \bise. We take the [0001] \bise\ direction as $z$ and
the (111) plane as the $xy$ plane. Along the $z$ direction the
supercells contain the Cr film on top
of either 4 or 6 QLs (20 or 30 atomic planes) of \bise\ and a vacuum layer 
larger than 20 \AA\ to avoid interaction between
opposite surfaces. During the structure optimization,
 the Cr overlayers and the interfacemost QL of \bise\ 
  were fully relaxed until the
  residual forces were smaller than 0.02~eV/\AA, while the remaining atoms were fixed 
  to the relaxed geometry of the corresponding \bise\ thin film.

\section{Atomic Structure and interface charge transfer}
\begin{figure*}
 \includegraphics[scale=0.17]{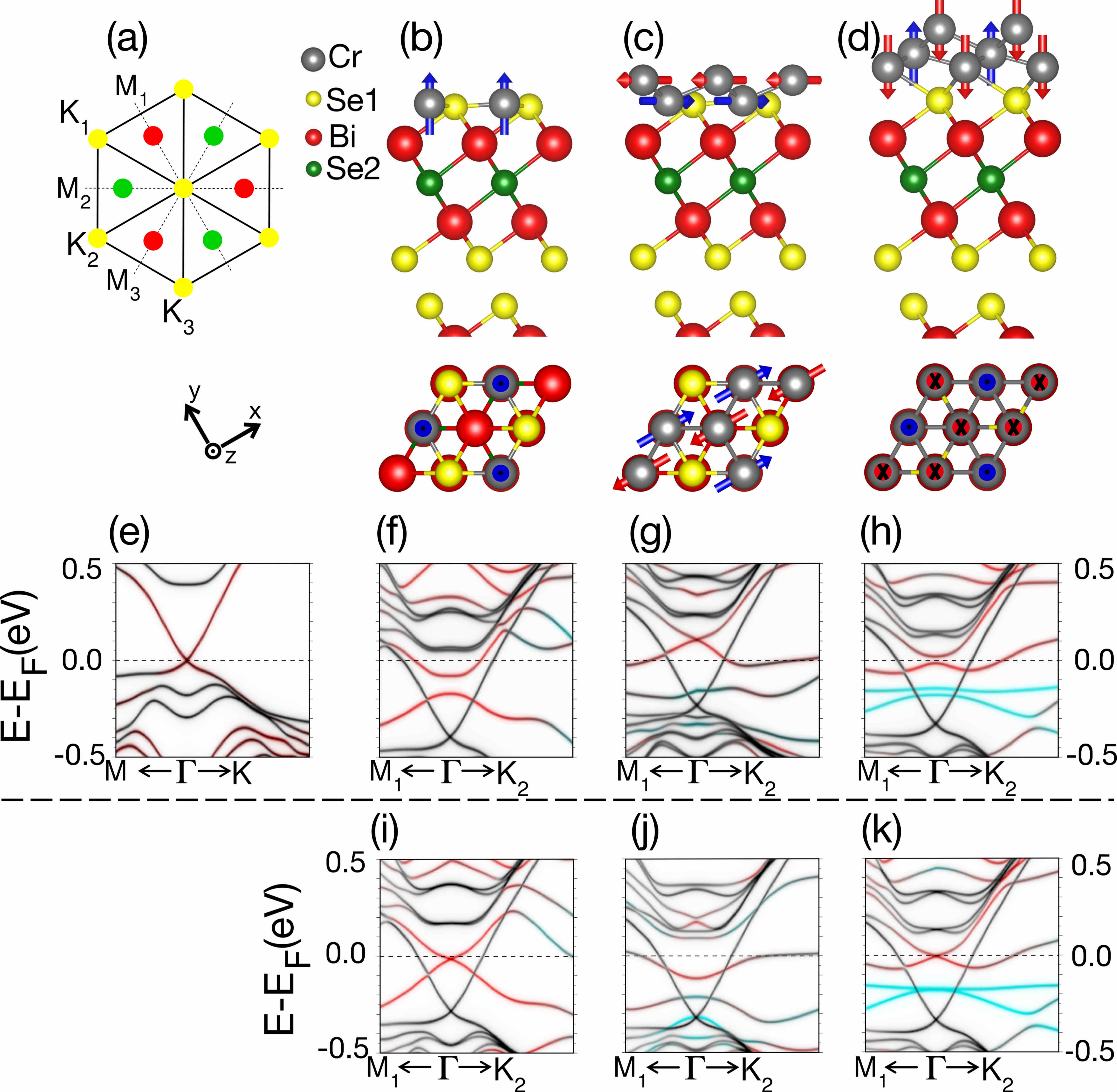}
\caption {(a) Top view of \bise\ (111) surface. Dashed lines depict the mirror planes M$_1$, M$_2$ and M$_3$, 
         and the M and K points of the Brillouin zone are also indicated. (b) to (d) show the relaxed
         geometries for 1 to 3 ML Cr coverages on a 4~QL \bise\ slab, along with arrows indicating the magnetization of the Cr layers
         for the magnetic ground state in each case.
         (e) Band structure around the center of the Brillouin zone for a pristine 4~QL \bise\ slab. (f) to (h) display 
         the band dispersion diagrams for 1 to 3 Cr MLs on a 4~QL \bise\ slab in the magnetic
	 ground state configuration shown above. (i) to (k) Depict the band structure of 1 to 3 Cr MLs on a 4~QL \bise\ 
	 slab with the Cr MMs perpendicular to that of the magnetic ground state for each system, \textit{i.e.} 
	 along $x$ for (i) and (k) and along $z$ for (j). The projection of the 
         states on the interfacemost QL of \bise\ is shown in red, while the projection on the Cr subsystem is shown in cyan.}
         \label{geom} 
\end{figure*}

\begin{figure}
 \includegraphics[scale=0.115]{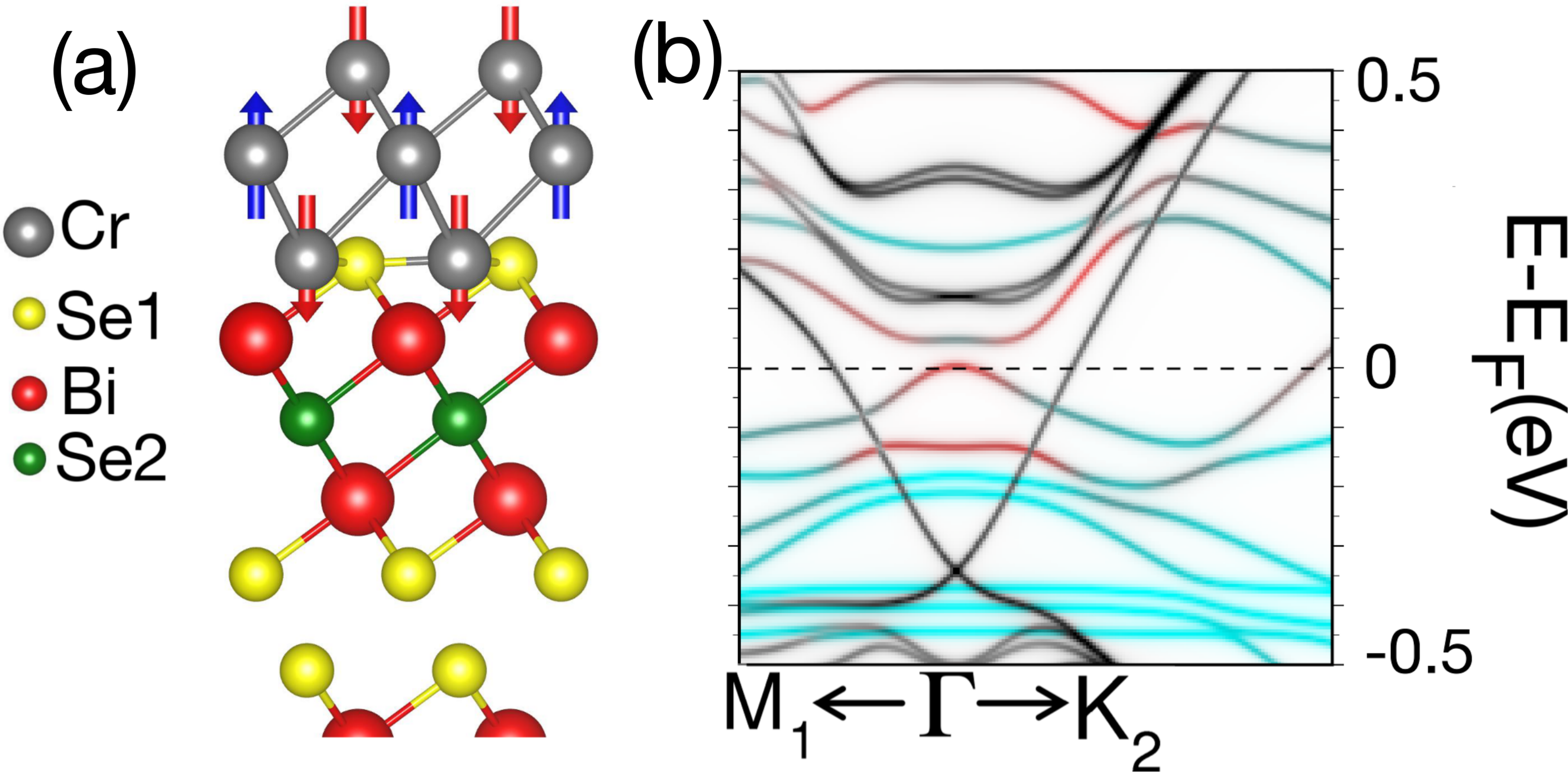}
\caption {(a) Side view of the 3 Cr ML on the \bise\ (111) surface with the Cr layers following the stacking 
         pattern of \bise, i.e. with the first Cr layer occupying the \textit{fcc} hollow site.
	 The arrows indicate the magnetic ground state for this ionic configuration. Note that the ionic configuration shown 
	 in Fig.~\ref{geom} (d) is more stable for the Cr trilayer.
	 The band structure of a 4 QL \bise\ slab with a Cr trilayer in the ionic and magnetic configuration depicted in (a)
	 is shown in (b). The projection of the 
         states on the interfacemost QL of \bise\ is shown in red, while the projection on the Cr subsystem is shown in cyan.}
         \label{fcc} 
\end{figure}

We consider commensurate Cr films with 1, 2 and 3 ML thicknesses on top
of (111) \bise\ surfaces. 
The atomic structure of the (111) composed slab exhibits three-fold
C$_{3}$ symmetry and three reflection planes perpendicular to the surface --see Fig.~\ref{geom} (a)--.
The in--plane lattice parameters of \bise\ (4.14~\AA) and Cr (4.12~\AA)
 show a small lattice mismatch of 0.5\%.
 First, we examine different positions 
for the Cr overlayers, including \textit{fcc} and \textit{hcp} hollow sites, bridge and Se-top sites. As expected,
the high symmetry hollow sites are the energetically most favorable. 
Figures~\ref{geom} (a) to (d) show 
the calculated equilibrium structures.
For 1 and 2 ML 
films the interfacial Cr atoms occupy the \textit{fcc} hollow 
sites following the \bise\ stacking, ...-BcAbC-\textbf{A} and 
...-BcAbC-\textbf{AB} respectively, where bold letters correspond to Cr atoms.
However, for the 3 ML film the interface Cr
moves into the \textit{hcp} hollow site on top of the Bi subsurface layer and there is a reversal 
of the stacking sequence~\cite{tunabledirac}, ...-BcAbC-\textbf{BAC}. 
This spatial
self--organization of the Cr film has to be due to the peculiar open structure of the (111)
\textit{bcc} surface in which first n.n.s are in the adjacent layers and 
in the third layers above and below. In this way, while for the 1 and 2 ML Cr films the interface Cr atoms are 
almost coplanar to the surface Se and lie on top of 
the Se in the center of the first QL, for the 3 ML film the Cr-Se interface bond 
distance increases notably and the Cr at the interface lies on top of 
the outermost Bi.
The relaxed bond lengths are given in Table~\ref{disttable}. The Cr-Cr distances  are
close to
the bond lengths in bulk Cr, 2.49~\AA. Note the increase in the Cr1-Se1 bond length 
for the 3 ML film. 
Additionally, the bond distances for the non-equilibrium Cr
 trilayer in the \textit{fcc} configuration --see Fig.~\ref{fcc} (a)-- are 
presented at the bottom of the Table. In this configuration, similar to the 1 and 2 ML cases, 
the interface Cr atoms remain almost coplanar to the Se surface at the expense of very large n.n.s Cr-Cr bond distances. The \textit{fcc} configuration is about 
80~meV more energetic than the equilibrium 3 ML Cr-\bise\ structure, well above the energy involved in room temperature fluctuations.

The calculated binding energies are also
given in the Table. The binding energy $E_\mathrm{ads}$ is obtained as
\begin{equation}
E_\mathrm{ads}=E_{\mathrm{Cr-Bi}_2\mathrm{Se}_3}-E_{\mathrm{Bi}_2\mathrm{Se}_3}-E_\mathrm{Cr}
\end{equation}
where $E_{\mathrm{Cr-Bi}_2\mathrm{Se}_3}$ is the total energy for the composed Cr--\bise\ system,
 $E_{\mathrm{Bi}_2\mathrm{Se}_3}$ is the total energy of the isolated 4 QL \bise\ system and 
$E_\mathrm{Cr}$ is the total energy of the isolated Cr subsystem in the same ionic and magnetic 
configuration as it acquires in the composed Cr-\bise\ system.
We found a negative value for the adhesion 
energy for all the Cr films in correspondence with the exothermic character of dilute Cr
adsorbed on \bise\ surfaces for submonolayer coverages~\cite{PhysRevB.88.045312,nanoCr}. 

\begin{table}
\centering
 \begin{tabular}{cccc|c}
            & Cr3-Cr2 & Cr2-Cr1 & Cr1-Se1&$E_\mathrm{ads}$   \\ \hline
  1 ML      & -       & -       & 2.39   &-2.00               \\
  2 ML      & -       & 2.59    & 2.40   &-1.98                \\
  3 ML      & 2.64    & 2.48    & 2.84   &-1.77                 \\
  3 ML-fcc* & 3.11    & 3.42    & 2.39   &-1.69                 \\
 \end{tabular}
 \caption
{Relaxed bond lengths in \AA\ between the Cr layers --columns 2 and 3-- and between the interface Cr 
 and the interface Se --column 4--. 
 The last row corresponds to the more energetic \textit{fcc} configuration (see
 Fig.~\ref{fcc}) for the 3 ML Cr system.
  The adhesion energy $E_\mathrm{ads}$ is given in eV in the rightmost column.
 The \textit{fcc}--like case for the 3 ML Cr is more than 80~meV less stable. 
}
 \label{disttable}
\end{table}

\begin{figure}
  \includegraphics[scale=0.5]{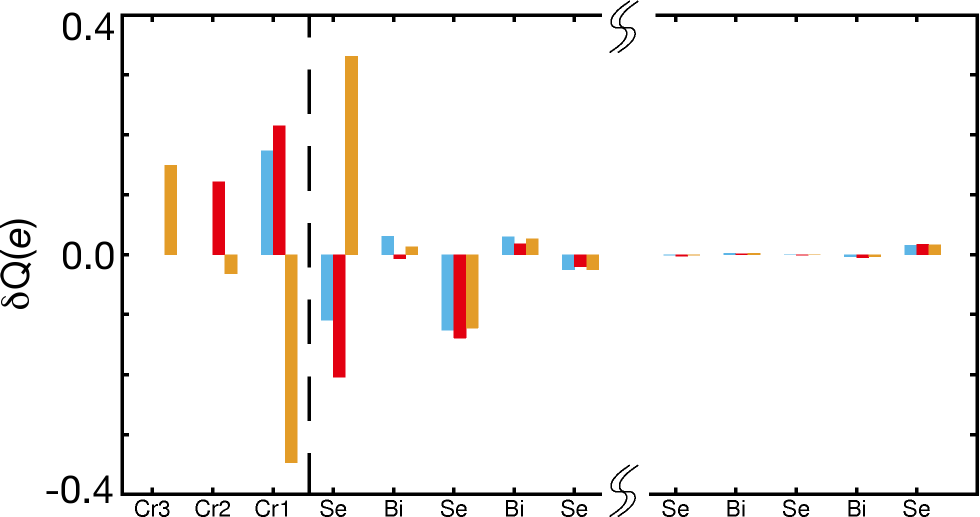}
  \caption{ Mulliken charge rearrangement of the 1 (cyan), 2 (red) and 3 (orange) Cr ML systems on a 4 QL \bise\ slab.
            The figure displays the atomic charge difference between the isolated Cr and \bise\ subsystems and that 
            acquired in the composed Cr--\bise\ system.
            Only the Cr subsystem and the first and last QLs are shown since the charge rearrangement in the inner QLs
            of \bise\ is negligible. The Cr--Se interface is indicated with a dashed line as a guide to the eye.}
  \label{mulliken}
 \end{figure}

The different atomic configuration of the equilibrium structures is clearly 
reflected in the interface charge redistribution. We have calculated the 
Mulliken charges for the Cr--\bise\ systems and for the corresponding isolated  
slabs, a pristine 4 QL \bise\ slab, and 
 the isolated Cr films of 1, 2 and 3 Cr MLs with the same atomic and 
 magnetic configuration as they present when adhered to \bise\ .
 The differences between the Mulliken charges of the entire Cr--\bise\ systems
and those corresponding to the isolated subsystems are displayed in Figure~\ref{mulliken}. 
In all the cases the charge transfer is small and mostly confined to the Cr film and the first \bise\ QL. For 1 and 2 Cr ML coverages, the Cr layers acquire charge
 at the expense of the Se atoms, both at the interface and in the middle of the first
QL. 
In the 3 Cr ML system, on the contrary, the charge transfer is towards the \bise\ .
The interfacial 
 Cr donates charge, mainly to the n.n.s. Se, which gains electron charge, 
increasing its ionic radius and consequently increasing the interface bond length. This different behavior can be attributed to the 
 different adsorption site of the first Cr layer (\textit{hcp} hollow versus \textit{fcc} hollow for 1 and 2 Cr MLs).
Nevertheless, there is always a chemical interaction at the interface. In addition, the \bise\
free surface presents a small charge gain in all the calculated structures.

\begin{figure}
 \includegraphics[scale=0.38]{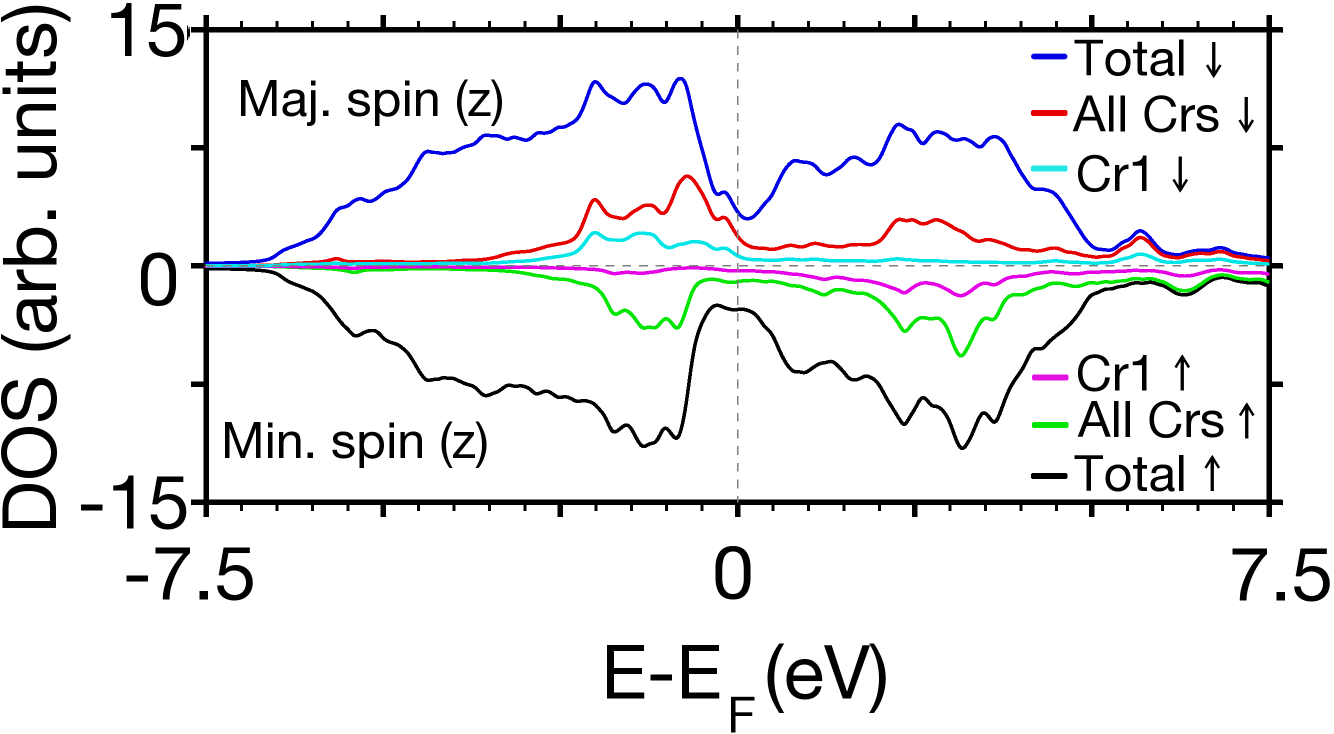}
\caption{Spin--resolved DOS for the Cr trilayer adhered to \bise.
 The blue (black) line shows the majority (minority) spin total DOS, while
 the red (line) corresponds to the DOS projected on the whole Cr trilayer
 for the majority
 (minority) spin bands and the cyan (magenta) indicates the DOS projected
 on the interface--most Cr layer for the majority (minority) spin. Magnetism
 is patent from the difference in the majority and
 minority curves.}
 \label{dos}
\end{figure}

\section{Magnetic ground state}

To model the magnetic ground state of Cr layers we consider different  
configurations having parallel and antiparallel collinear Cr magnetizations both between 
planes and within a plane. We employed an in--plane unit cell with 
2 atoms per plane. We find a ferrimagnetic ground 
state with ferromagnetic Cr planes coupled antiferromagnetically for all the studied Cr film thicknesses. 

Due to the C$_{3}$ symmetry of both Cr and \bise\ layers,
the 3$d$ Cr and 4$p$ Se orbitals 
hybridize, as can be clearly appreciated in
the spin--resolved total DOS for three Cr layers adhered to \bise\ shown in Fig.~\ref{dos}. The hybridization
drives the Cr states close to
the Fermi level, confined in an energy region $\approx$ 1.5~eV below~E$_F$.
In addition, a large energy splitting of about 4~eV between the spin--majority 
and the spin--minority states is obtained, and the majority Cr states  
are fully occupied while the minority--spin channel is almost unoccupied.  
Therefore, the  magnetic moments (MMs) of the Cr layers are close to the Hund rule value for isolated Cr atoms.
The calculated MMs, shown in Table~\ref{MMtable}, are remarkably large 
at the surface plane ($\geq$ 4 $\mu_B$/atom) for all the systems, while 
they decrease for the subsurface Cr layers.
For 1 and 2~ML Cr films there is an 
appreciable induced MM on the Se and Bi topmost planes of 0.2~$\mu_B$, aligned
opposite to the Cr MM at the interface, while the induced MMs in the \bise\ for the
3 ML Cr film is almost negligible in correspondence with the different chemical interaction at the interface.
Note the larger MMs of the 3 ML Cr \textit{fcc} structure due to larger interlayer distances.

\begin{table}
\centering
 \begin{tabular}{cccc|c}
           &$\mu_{\mathrm{Cr3}}$&$\mu_{\mathrm{Cr2}}$&$\mu_{\mathrm{Cr1}}$&$\mu_{\mathrm{Tot}}$ \\ \hline
  1 ML     &  -                 &  -                 & 4.3                & 4.0                  \\
  2 ML     & -                  & 4.2                &-3.1                & 1.4                   \\
  3 ML     &-4.2                & 3.6                &-3.7                &-4.3                    \\
  3 ML-fcc*&-4.9                & 4.7                &-4.0                &-3.8                    \\
 \end{tabular}
 \caption
{Magnetic moments of the Cr layers in Bohr magnetons for 1, 2 and 3 ML coverages. Cr1 (Cr3)
 corresponds to the interfacemost (farthest from the interface) Cr layer. $\mu_{\mathrm{Tot}}$ 
is the total magnetization of the whole Cr--\bise\ system for each case.
 The last row corresponds to the more energetic \textit{fcc} configuration (see
 Fig.~\ref{fcc}) for the 3 Cr ML system.
 Cr overlayers grow as a layer--by--layer ferrimagnet with in--plane ferromagnetic coupling.
}
 \label{MMtable}
\end{table}

Since the spin--orbit coupling is included in the calculations we can determine
the direction of the Cr MM relative to the crystal lattice.
The preferential orientation of the Cr magnetization vector was obtained by
comparing the total energies of in--plane ($\mathcal{M}_x$, $\mathcal{M}_y$) and
out--of--plane ($\mathcal{M}_z$) orientations of the total 
magnetization $\bb{\mathcal{M}}$ (the $z$ axis is defined normal to the surface). It is noteworthy to point out that in the ground state
within the planes the Cr atoms are always coupled ferromagnetically, thus the
Cr MMs are aligned within each layer--see Fig.~\ref{geom} (a)--. 

The easy magnetization axis for the 1 ML Cr--\bise\ system lies perpendicular to the surface
(out--of--plane), while as the thickness of the Cr film increases a double spin 
reorientation transition takes place and the magnetization direction changes
to in--plane for 2 ML and again to out--of--plane for the 3 ML Cr film. A similar
spin reorientation transition has been reported in ultrathin Co films grown on 
hexagonal Ru (0001)~\cite{CoRu}. The magnetic anisotropy for the 1 and 2~ML Cr--\bise\ systems
 is unusually large, of $\approx$~25~meV and 35~meV respectively, while for the 3~ML Cr--\bise\ system
 we obtain a smaller value of 5~meV.

\section{Topological surface states} 
We additionally analyze the electronic structure of the Cr--\bise\ slabs. Figures~\ref{geom} (e) to (h)
 show the corresponding band dispersions around the $\Gamma$ point and that 
of the pristine \bise\ 4~QL film. The band dispersion of the 
pristine film shows the topologically protected metallic surface states with the 
Fermi level located at the Dirac point. However, for all the Cr--\bise\ 
slabs the position of the Fermi level is shifted up between 0.2 and 0.4~eV 
with respect to the Dirac cone of the free \bise\ 
surface, which persists in the three cases. 
As a result the free surface topological SSs are always electron doped.
 
Next, we focus on the SS when a single Cr overlayer is 
adhered to the \bise\ surface. A large Dirac gap opens up, and the gap 
opening only occurs at the interface with the magnetic film while the Dirac cone at the free
\bise\ surface remains, evidencing the 
spatially localized character of the effect. Furthermore, our calculations reveal that the magnetic 
easy axis is along the out--of--plane direction as shown in Fig.~\ref{geom} (b). 
Therefore, the origin of the gapped Dirac 
point is the exchange coupling between the TI SS and the 
out--of--plane magnetization of the Cr film, which breaks TRS.

As explained above, in the 2 ML system the Cr layers present an
in--plane magnetization, and we do not find any appreciable energy difference 
when the in--plane magnetization is along or normal to the vertical reflection planes
of the \bise\ thin films --see Figure~\ref{geom} (a)--. Thus, we discuss the results for the in--plane 
magnetization normal to the reflection plane M$_1$. The corresponding
band dispersion around the $\Gamma$ point is represented in Figure~\ref{geom} (g). 
The topological surface state survives and there is no shift in momentum 
space of the Dirac point, which remains at $\Gamma$. However, the dispersion
 is no longer linear and the SS presents a large anisotropic mass. Only along the 
-K$_2$--$\Gamma$--K$_2$ line, perpendicular to the mirror plane,
 electrons at $\bb{k}$ and  $-\bb{k}$ have the same 
energy.
The preservation
of the Dirac point can be easily understood considering that although the 
breaking of TRS occurs for any non--zero magnetization, 
the slab is invariant under a reflection normal to the in--plane magnetization 
direction, thus the reflection symmetry M$_1$ survives. 
This
result is a clear demonstration that in order to open a gap at the Dirac cone,
breaking the TRS and the three reflection symmetries M$_{1,2,3}$ of the \bise\ 
lattice is required~\cite{InplanePRL}.
As in the 1 ML system, the Dirac cone at the free surface of \bise\
remains unmodified but for an energy shift. 

For the system consisting of 3~MLs of Cr on top of the \bise\ thin film,
the magnetization points again along the out--of--plane direction. Therefore, its behavior is 
analogous to that of the 1 Cr ML slab: a gap opens at the original Dirac point, 
although  the gap is smaller. Moreover, it is worth to note that for 3 Cr MLs, 
the Fermi level lies exactly within the gap of the surface Dirac fermions 
gapped by the exchange interaction.

For comparison, we have additionally included the dispersion relations of
the 1, 2 and 3 Cr-\bise\ systems
with the magnetization of the Cr layers aligned perpendicular to that of the 
corresponding magnetic ground states, \textit{i.e.} in--plane along $x$ for
the 1 and 3 ML Cr and out--of--plane along $z$ for the 2 ML Cr case
--Fig.~\ref{geom} (i) to (k)--.
 Now, the behavior of the topological SS is just
the opposite, which confirms the correlation between the opening of the gap at the 
Dirac point and the presence of a perturbation that breaks both TRS and
the invariance of the system under the three reflection symmetries of the \bise\ 
lattice. The crossing of the topological SS persists whenever the magnetization
is aligned in--plane and perpendicular to a reflection plane, as in the
1 and 3 Cr ML systems --Fig.~\ref{geom} (i) and (k)--. In both cases the reflection symmetry M$_1$ is 
preserved. On the contrary, a gap opens for the out--of--plane 2 ML Cr film, where TRS 
and the three reflection symmetries M$_{1,2,3}$ are broken. The mass 
enhancement and the induced anisotropy in the topological SS for the 1 and 3 Cr MLs
are also clearly appreciable. Moreover, the origin of the large calculated 
MAE is evident from the sharp contrast between 
the band structures of these excited states -- Fig.~\ref{geom} (i) to (k)-- 
and their corresponding magnetic ground states -- Fig.~\ref{geom} (f) to (h)--.
Finally, the band structure of the non-equilibrium 3 ML Cr film with the 
\textit{fcc} stacking is shown in Fig.~\ref{fcc} (b). As expected, there is a gap 
opening due to the out--of--plane magnetization, analogous to that developed in
the equilibrium 3 ML Cr-\bise\ structure --see Fig.~\ref{geom} (h)--.

These results prove that the gap opening of the topological surface states is exclusively due to the interplay
of the topology
and the induced magnetization, and independent of the chemical behavior.
As remarked above the 1 and 2 ML Cr-\bise\ systems show similar interface chemical 
interactions --the charge transfer has the same sign and similar value-- and opposite
to the interface interaction in the 3 ML slab (see Figure~\ref{mulliken}).
Nevertheless, there is a gap in the 1 and 3 ML Cr-\bise\ systems, while in the 
2 ML Cr-\bise\ structure the degeneracy of the topological SS at the $\Gamma$ point remains.

\begin{figure*}
\includegraphics[scale=0.28]{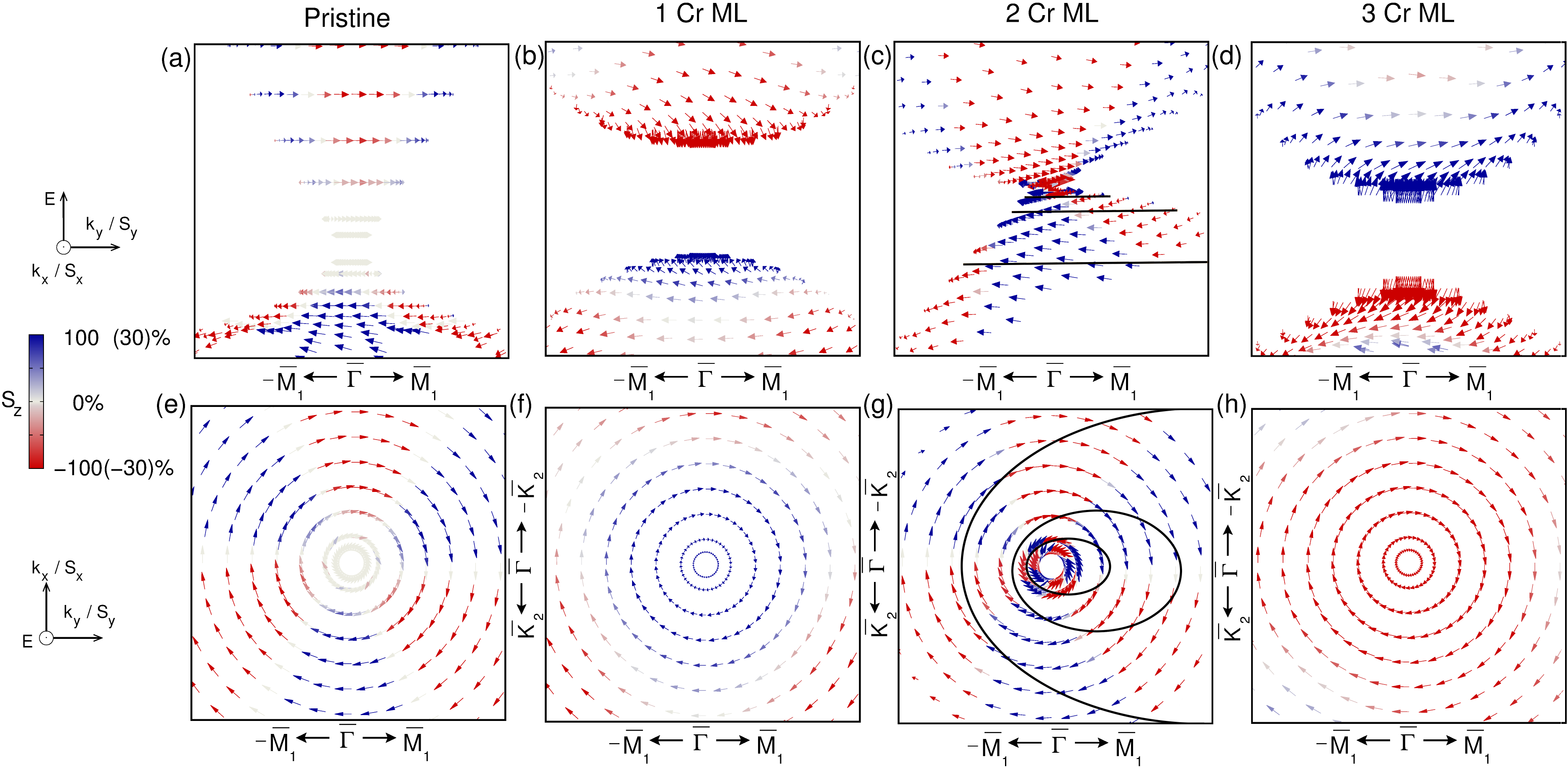}
\caption{(a) to (d): Side view of the spin texture of the surface state for Cr overlayers of 0 (pristine \bise\ surface) 
         to 3 MLs. (e) to (h): Top view of the hole--like surface state for Cr coverages of 0 to 3 MLs.
         The expectation value for the spin is shown as an arrow at each $k$--point, while the S$_z$ component is additionally
         color coded according to the scale at the right, being the limits $\pm$100 (30)\% of
         the modulus of S=$\sqrt{\mathrm{S}_x^2+\mathrm{S}_y^2+\mathrm{S}_z^2}$
         for the 1 and 3 (0 and 2) Cr MLs. In the 2 Cr ML system --(c) and (g)-- three elliptical black solid lines depict constant energy
	 contours. The circular meron texture is patent in the 1 and 3 ML 
         cases, while the spin texture of the 2 Cr MLs on \bise\
         is an anisotropic circular skyrmion.}
         \label{spin} 
\end{figure*}

\section{Spin Texture of the Surface States}
As shown above, the magnetization of the Cr layers attached to the surface of 
the \bise\ film provides a local magnetic field, which modifies the 
degeneracy and topology of the SS. Additionally, it induces a spin 
component along the magnetization direction and alters the spin texture of the 
topological SS. We examine the spin texture of the SSs in the equilibrium Cr--\bise\ systems close 
to $\Gamma$ by calculating the expected value of the spin operator. 
The results are displayed in Figure~\ref{spin}, which also includes the spin distribution
of the Dirac cone states of the pristine \bise\ surface. For the latter
the spin is locked perpendicular to crystal momentum, showing the distinct helical 
spin texture protected by TRS, and S$_z$ vanishes close to the Dirac point. At large $k$
there is, however, a finite small S$_z$ component due to the trigonal warping. S$_z$ remains null
along the mirror lines $\Gamma$-M and 
reverses its sign traversing from K to -K, in correspondence with the trigonal symmetry of the system.

The spin texture of the gapped topological SSs (1 and 3 ML Cr 
systems) is in sharp contrast to that  of
the free surface. In the vicinity of the gapped Dirac point, the states show 
an imbalance between
S$_z$  and -S$_z$ at a given energy, and they present a significant net out--of--plane spin 
polarization. Only the in--plane components reverse sign changing from $\bb{k}$ to -$\bb{k}$.
Furthermore, the upper and lower Dirac bands 
have opposite S$_z$, evidencing that the spin degeneracy is indeed lifted 
at the $\Gamma$ point.
For larger $\bb{k}$, away from $\Gamma$, the induced S$_z$ component gradually
decreases, and the out--of--plane spin distribution results from the competition between the 
magnetic order that aligns the spin along the out--of--plane direction and the
spin texture imposed by the warping term which forces adjacent K points
to have opposite S$_z$. 
In the 2 ML Cr slab, the in--plane magnetization exhibited by the Cr layers 
in the interfacial plane does not induce observable spin reorientations of the Dirac
 state, and its spin texture is analogous to that of the free surface Dirac cone. However, 
due to the large anisotropy of the 
effective mass, the
constant energy lines are no longer circular, but present an elliptical shape. 
Nevertheless, the SSs exhibit a well defined spin helicity and the total 
spin cancels in every constant energy contour.
TRS breaking is evident from the spin texture of the three Cr--\bise\ systems analyzed.

\section{Conclusions} 
In summary, we have found that the structural configuration of 
ultrathin Cr films
attached to the (111) surface of \bise\ is determinant to establish
the topological behavior of \bise\ SSs. Due to the coupling 
between Cr 3$d$ orbitals and the \bise\ electrons, the Cr interface 
induces simultaneous charge and magnetic doping. However, the properties of the
topological SS critically depend on the Cr film thickness and are independent
of the specific chemical interaction at the Cr-\bise\ interface. As the thickness
of the Cr film increases stepwise from one to three MLs, the 
magnetization of the Cr layers undergoes two reorientation transitions, 
and changes from out--of--plane (1 ML) to in--plane (2 ML) and to
out--of--plane (3 ML) once again. For the 1 ML and 3 ML Cr--\bise\ interfaces 
the magnetic overlayer induces a gap at the Dirac point,
producing massive fermions at the interface.
 Moreover, the gap already opens for a single Cr ML, and
the value of the gap depends on the absolute value of the exchange interaction. 
In contrast, for the 2 ML Cr 
system the gapless Dirac cone is preserved. The complexity of 
the spin texture of gapped Dirac states signifies a competition between
the in--plane helical component of the spin dictated by the spin--orbit 
coupling and the out--of--plane TRS 
breaking component induced by the proximity to the magnetic Cr. Our results evidence the importance of the actual 
structural configuration of the magnetic films and show that the thickness of the Cr film can be used to modify
in a controlled way the metallic or gapped nature of topological Dirac states and their associated spin texture. 
\section*{Acknowledgments}
This work has been supported by the Spanish Ministry of Economy and Competitiveness
 through Grants MAT2012-38045-C04-04 and MAT2015-66888-C3-1-R. We acknowledge the use of computational resources of 
 CESGA, Red Espa\~nola de Supercomputaci\'on (RES) and the i2BASQUE academic 
 network. We also acknowledge J.I. Cerd\'a for fruitful discussions.
\bibliography{Bibliolinks}
\end{document}